\documentclass[aps,prd,reprint,preprintnumbers,amsmath,amsfonts,amssymb]{revtex4-2}

\usepackage{graphicx}
\usepackage{epsfig}
\usepackage{bm}
\usepackage{hyperref}
\hypersetup{colorlinks,linkcolor=blue,citecolor=blue,urlcolor=blue,breaklinks}
\usepackage[hyphenbreaks]{breakurl}
\usepackage{listings}
\usepackage{xcolor}
\usepackage{DejaVuSansMono}

\newcommand{\mono}[1]{\tt{}\footnotesize#1\normalsize\rm{}}

\definecolor{ansiwhite}{rgb}{0.95, 0.95, 0.95}
\definecolor{ansimagenta}{rgb}{0.9, 0.0, 0.9}
\definecolor{ansigreen}{rgb}{0.0, 0.85, 0.0}
\definecolor{ansired}{rgb}{0.9, 0.0, 0.0}

\lstdefinestyle{mystyle}{
    basicstyle=\ttfamily\footnotesize,
    backgroundcolor=\color{ansiwhite},
    commentstyle=\color{ansigreen},
    keywordstyle=\color{ansimagenta},
    numberstyle=\color{ansigreen},
    stringstyle=\color{ansired},
    captionpos=b,
    numbers=left,
    numbersep=5pt,
    showstringspaces=false,
}

\begin{document}

\title{Use QUDA for lattice QCD calculation with Python}

\author{Xiangyu Jiang}
\email{jiangxiangyu@itp.ac.cn}
\affiliation{Institute of High Energy Physics, Chinese Academy of Sciences, Beijing 100049, People's Republic of China}
\affiliation{Institute of Theoretical Physics, Chinese Academy of Sciences, Beijing 100190, People's Republic of China}

\author{Chunjiang Shi}
\affiliation{Institute of High Energy Physics, Chinese Academy of Sciences, Beijing 100049, People's Republic of China}

\author{Ying Chen}
\affiliation{Institute of High Energy Physics, Chinese Academy of Sciences, Beijing 100049, People's Republic of China}

\author{Yi-Bo Yang}
\affiliation{Institute of Theoretical Physics, Chinese Academy of Sciences, Beijing 100190, People's Republic of China}
\affiliation{School of Physical Sciences, University of Chinese Academy of Sciences, Beijing 100049, People's Republic of China}
\affiliation{School of Fundamental Physics and Mathematical Sciences, Hangzhou Institute for Advanced Study, UCAS, Hangzhou 310024, People's Republic of China}
\affiliation{International Centre for Theoretical Physics Asia-Pacific, Beijing/Hangzhou, People's Republic of China}

\author{Ming Gong}
\affiliation{Institute of High Energy Physics, Chinese Academy of Sciences, Beijing 100049, People's Republic of China}

\begin{abstract}
  We developed PyQUDA\footnote{Git repository: \url{https://github.com/CLQCD/PyQUDA}}, a Python wrapper for QUDA written in Cython, designed to facilitate lattice QCD calculations using the Python programming language. PyQUDA leverages the optimized linear algebra capabilities of NumPy/CuPy/PyTorch, along with the highly optimized lattice QCD operations provided by QUDA to accelerate research. This integration simplifies the process of writing calculation codes, enabling researchers to build more complex Python packages like EasyDistillation for specific physics objectives. PyQUDA supports a range of lattice QCD operations, including hybrid Monte Carlo (HMC) with N-flavor clover/HISQ fermions and inversion for the Wilson/clover/HISQ fermion action with the multigrid solver. It also includes utility functions for reading lattice QCD data stored in Chroma, MILC, and $\chi$QCD formats. Type hints are supported by stub files and multi-GPU support is provided through mpi4py.
\end{abstract}

\maketitle

\section{Introduction}\label{section:intro}

QUDA~\cite{Clark:2009wm,Babich:2011np} is a library for performing lattice QCD calculations on GPUs. It has been widely used in our recent work to accelerate inversion operations in the Chroma~\cite{Edwards:2004sx} framework. However, other operations, such as smearing in Chroma, are written using the QDP interface and implemented by QDP++~\cite{Edwards:2004sx} or QDP-JIT~\cite{Winter:2014dka}, which are less efficient compared to QUDA. Porting these operations to QUDA should make the overall calculations faster, but is not a easy work to do in Chroma. On the other hand, we have already used \mono{einsum} in NumPy~\cite{harris2020array}, CuPy~\cite{cupy_learningsys2017}, or PyTorch~\cite{Ansel_PyTorch_2_Faster_2024} to perform linear algebra operations, such as contractions for the two-point function, with great efficiency. This led us to the idea that the workflow would be more user-friendly if QUDA functions could be directly called from Python.

The basic motivation for introducing the PyQUDA package comes from the excellent \mono{einsum} function in CuPy. We had been using \mono{einsum} for a while in our workflow and wanted everything to happen within a single Python script. Binding QUDA's C API to Python seems like a good choice: it wraps the most useful operations in QUDA, allowing us to avoid numerous format conversions. Although we can use QUDA's native format and pass pointers directly to QUDA, we believe it is better to use a human-friendly format when we want to inspect specific values in the lattice field in Python. Additionally, some operations require manually setting wider borders for gauge fields, and the C API keeps these details away from us.

PyQUDA is a Python wrapper for QUDA written in Cython~\cite{behnel2010cython}, dedicated to providing a user-friendly Python interface that allows researchers to call high-performance QUDA functions directly from Python. Data are stored as NumPy or CuPy NDArray, making it easy to use general scientific computing functions in NumPy or CuPy for data processing and analysis. With PyQUDA, the entire workflow---from gauge configuration generation to quark propagator computation to data analysis---can be implemented in Python, which is particularly convenient for small-scale proof-of-concept cases. We have also made the wrapper as thin as possible to maximize QUDA's performance, and the efficiency appears to be quite good in production. PyQUDA ships stub files containing type hints~\footnote{{PEP 484 – Type Hints}, \url{https://peps.python.org/pep-0484/}} and enables code autocompletion through modern language server protocol~\footnote{{Official page for Language Server Protocol}, \url{https://microsoft.github.io/language-server-protocol/}} (LSP) or integrated development environment (IDE). PyQUDA is open source under the MIT license.

Lyncs-API~\cite{Bacchio:2022bjk} is another project that enables efficient lattice QCD calculations with Python and also provides access to QUDA functions~\cite{Yamamoto:2022ygt}. It exposes QUDA's C++ API to Python, allowing class instances from QUDA to be easily managed by Python objects. This is a convenient way to call numerous C++ functions in QUDA. Lyncs-API uses cppyy package~\cite{7836841} to dynamically bind C++ functions to Python. The main difference between our work and Lyncs-API is that we use Cython to statically bind QUDA's C API to Python, which is more stable and offers a little better performance when calling bound functions~\footnote{{Benchmarks - nanobind documentation}, \url{https://nanobind.readthedocs.io/en/latest/benchmark.html}}. Grid Python Toolkit (GPT)~\footnote{{lehner/gpt: A Python toolkit for lattice field theory, quantum computing, and machine learning}, \url{https://github.com/lehner/gpt}} is the Python interface for the Grid~\cite{Boyle:2015tjk} project, which also supports GPU calculations for lattice QCD. It's possible to communicate with these packages as we are using NumPy's NDArray as the basic format to store the data.

PyQUDA has some prerequisites for building and performing calculations, which we list below:
\begin{itemize}
  \item pycparser~\footnote{{eliben/pycparser: :snake: Complete C99 parser in pure Python}, \url{https://github.com/eliben/pycparser}} is a parser for the C language, written in pure Python.
  \item Cython is a Python compiler that makes writing C extensions for Python as easy as Python itself.
  \item NumPy is the fundamental package for scientific computing with Python.
  \item CuPy is a NumPy/SciPy-compatible array library for GPU-accelerated computing with Python. (optional)
  \item PyTorch is a Python package that provides tensor computation (like NumPy) with strong GPU acceleration and deep neural networks built on a tape-based autograd system. (optional)
  \item MPI for Python (mpi4py)~\cite{Dalcin_mpi4py_Status_Update_2021} package provides Python bindings for the Message Passing Interface (MPI) standard.
\end{itemize}
First, we use the pycparser package to parse the QUDA C API header file \mono{quda.h}. A script then converts all struct and enum declarations in the header to Cython header and source code. We need to manually write the definitions of these functions in Cython to pass pointers to QUDA. PyQUDA is designed to be compatible with recent versions of QUDA, as these function declarations have changed recently. Once we have the Cython source code, we can use Python's setuptools and wheel packages to build and install the package.

All descriptions of PyQUDA in this article are based on version 0.9.5, which is currently in the early alpha stage, and the API may change. The article is organized as follows: Section~\ref{section:impl} describes the implementation details of the package; Section~\ref{section:propag} provides an example of calculating the quark propagator using the anisotropic clover fermion action~\cite{Sheikholeslami:1985ij,Klassen:1998fh} in Python along with the method to enable the multigrid solver in QUDA~\cite{Clark:2016rdz}; and Section~\ref{section:hmc} presents another example involving rational hybrid Monte Carlo (RHMC)~\cite{Clark:2006wq} with the Symanzik gauge action~\cite{Weisz:1982zw} and $N_f=2+1$ isotropic clover fermion action\cite{Sheikholeslami:1985ij}, in which tadpole improvements~\cite{Lepage:1992xa} are used. Finally, Section~\ref{section:summary} contains the summary and the outlook.

\section{Implementation\label{section:impl}}

\subsection{Convention\label{subsection:convention}}

We follow Chroma's convention in most of our work. We use the DeGrand-Rossi basis for the Dirac matrices and the QDP convention for our field ordering. Consider a 4D ($N_d=4$) lattice with size $[L_x, L_y, L_z, L_t]$, number of spins $N_s$ and number of colors $N_c$, we can define several field types on the lattice. The lattice layout in the QDP convention (with cb2 layout) is even-odd preconditioned, which is $[2, L_t, L_z, L_y, L_x/2]$ with even-odd parity order. The gauge field order is $[N_d, 2, L_t, L_z, L_y, L_x/2, N_c, N_c]$ with row-column color order, while the fermion field order is $[2, L_t, L_z, L_y, L_x/2, N_s, N_c]$. This format allows users to easily inspect specific values in these fields using Python and simplifies writing subscript strings in \mono{einsum}. Although QUDA needs to convert the data format to its native one on GPUs, this conversion typically occurs only at the beginning and end of a function, which doesn't take much time. We have not observed any noticeable performance regression using this convention.

The propagator field is not a direct input type of QUDA, but it is essential in lattice QCD research. The field order is $[2, L_t, L_z, L_y, L_x/2, N_s, N_s, N_c, N_c]$ with sink-source spin and sink-source color order. The clover field order is $[2, L_t, L_z, L_y, L_x/2, 2, (N_sN_c/2)^2]$ of real numbers, with upper-lower clover parity order, which also follows Chroma. The final $(N_sN_c/2)^2$ axes consist of diagonal (real) and off-diagonal (lower triangle, complex) parts of an anti-Hermitian matrix (the DeGrand-Rossi basis is one of the chiral bases). The momentum field used in the HMC algorithm is a combination of the QUDA convention and the Chroma convention. The field order is $[2, L_t, L_z, L_y, L_x/2, N_c^2/2+1]$ of real numbers with even-odd parity order. The last $N_c^2/2+1$ consists of the off-diagonal part (lower triangle, complex), diagonal part (real), and a conserved number (real).

However, we use the MILC convention~\footnote{{milc-qcd/milc\_qcd: MILC collaboration code for lattice QCD calculations}, \url{https://github.com/milc-qcd/milc_qcd}} with the staggered phase, which is defined as $\psi'(x,y,z,t)=\gamma_4^t\gamma_1^x\gamma_2^y\gamma_3^z\psi(x,y,z,t)$, where $\psi'$ is the fermion field with the staggered phase applied.
The Chroma convention defines the staggered phase as $\psi'(x,y,z,t)=\gamma_1^x\gamma_2^y\gamma_3^z\gamma_4^t\psi(x,y,z,t)$ and adds an extra minus to the off-diagonal part of the matrix. Chroma applies an SU(3) projection on the fat link created by the first fattening step in the highly improved staggered quark (HISQ)~\cite{Follana:2006rc} process, which differs from the U(3) projection in the MILC code. QUDA preserves the difference on the projection type. Since it is not a simple phase conversion between the results under the two conventions and we often use MILC-related code to calculate the HISQ propagator, it is more convenient to follow the MILC convention here. This can be easily changed by setting the parameter called \mono{staggered\_phase\_type} in \mono{QudaGaugeParam} if the Chroma convention is preferred.

All the data mentioned above are stored in corresponding Python classes named \mono{LatticeGauge}, \mono{LatticeFermion}, etc. Staggered fermion and propagator fields are also defined.

\subsection{Pointers with Python}

First, we need to work with C pointers in Python, which are very common in C functions. We defined a class named \mono{Pointer} with the C pointer as a member in the Cython source code, and the C pointer cannot be accessed from the Python side. This means we can only preserve but not create a pointer (except for a null pointer) in Python. Pointers are meaningful only after memory has been allocated, which corresponds to creating a NumPy NDArray in Python. We create a wrapper function \mono{ndarrayPointer} in Cython to extract the pointer from an existing NumPy NDArray. The function is compatible with CuPy's NDArray and PyTorch's Tensor, allowing us to get a pointer pointing to the GPU memory. The NDArray or Tensor stores the pointer as an integer, and explicitly casting the integer to a \mono{void *} gives us the pointer we need. All NDArray or Tensor objects should be contiguous in memory, and we ensure contiguity when creating and using all lattice fields.

On the other hand, an array of pointers is needed to handle the gauge field in the Chroma convention. We made the \mono{ndarrayPointer} wrapper compatible with NDArrays that have more than one dimension. The function creates an array of pointers and stores them in a \mono{Pointers} class. Note that NDArrays with more than three dimensions are not supported.

\subsection{Translate struct and enum}

An annoying aspect of Cython is that setter and getter functions must be defined explicitly to access members of a struct in C. This is acceptable in most situations but becomes problematic with QUDA's large parameter structs. There are hundreds of struct members, and we would have to manually write all these setter and getter functions in the Cython source code. To address this, we notice the similarity among these functions and use a script to translate these structs from C to Cython. Another issue with Python is that we must explicitly assign values to all enum names, whereas not all names in \mono{enum\_quda.h} have explicit values. We also use the script to translate enums from C to Python, assigning values to all names based on their implicit values in C.

With the help of the pycparser package, we can parse the \mono{quda.h} header along with \mono{quda\_constants.h} and \mono{enum\_quda.h} into an abstract syntax tree (AST). We pick all structs with names like \mono{QudaXxxxParam} and loop over all their members. These member variables can be classified into several categories according to their types: normal scalars (\mono{int}, \mono{double}), strings (\mono{char []}), arrays (\mono{int []}, \mono{double []}), arrays of strings (\mono{char [][]}), void pointers (\mono{void *}), scalar pointers (\mono{int *}, \mono{double *}), arrays of pointers (\mono{int **}, \mono{double **}), parameter pointers (\mono{QudaXxxxParam *}), and arrays of parameter pointers (\mono{QudaXxxxParam *[]}). We write templates for all these categories and generate the corresponding getter and setter functions for every struct member.

The translation of enums requires a counter for the value of a name. We find that some QUDA enum names have comments on the same line, so we decide to extract these comments and add them as docstrings in Python. This is a bit tricky and requires strong constraints on the QUDA enum comment format. This feature might break in the future but should be fixable with minor modifications.

Most of the binding can be generated automatically, but we manually translate functions in those header files to Cython source code. This approach prevents the package from working with every QUDA commit, as specific function declarations are needed. We are working to maintain compatibility with recent commits, and in the future, function translations will be generated automatically.

\subsection{Parallel computing with MPI}

In QUDA, the lattice are divided equally into several parts to enable multi-GPU computation. The partitioning is performed according to a 4D Cartesian grid, and the grid volume should be equal to the MPI size. The logic for assigning coordinates in the grid to different MPI ranks in PyQUDA is the same as QUDA's default. The rank varies fastest in the $t$ direction, followed by $z$, $y$, and $x$, which means all MPI processes are arranged as a 4D row-major array with size $[G_x, G_y, G_z, G_t]$. Each process will handle the corresponding part of the entire lattice. We apply even-odd preconditioning to all local field data on all processes. Note that the order of the field layout on a process described in Subsection~\ref{subsection:convention} ($[2, L_t, L_z, L_y, L_x/2]$) is different from that of the grid. The computation is performed on multiple GPUs by passing these pointers to QUDA.

Parallel I/O for some lattice file formats has been implemented. We define a subarray datatype in MPI I/O and then commit it. After setting the file view, we call the collective read function to read in data with good performance. We also wrote a simple parser for some complex formats like QIO~\footnote{{usqcd-software/qio: QIO Parallel IO Library}, \url{https://github.com/usqcd-software/qio}}. For example, we can read gauge configurations and quark propagators generated by Chroma and saved in QIO format. However, we cannot write data to a QIO file without introducing QIO itself. MILC files are supported for reading, and KYU and $\chi$QCD files are supported for both reading and writing. We also introduced an ``NPY'' format, which is the default file format for NumPy NDArray, making it easy for those familiar with NumPy to use the data.

\subsection{Python API}

After generating the binding files, we can now build them into a Python extension. We focus on creating a user-friendly interface, and Python greatly assists in this regard. Most of the work on the Python API involves the gauge matrix, the fermion matrix (or Dirac matrix) and the lattice field.

Let's start with the gauge matrix. First, we define a gauge matrix that contains a \mono{QudaGaugeParam}, a \mono{QudaInvertParam}, a \mono{QudaGaugeSmearParam}, and a \mono{QudaGaugeObservableParam}. It also handles precision and reconstruction settings. The gauge field properties can be set via \mono{QudaGaugeParam}, and some pure gauge operations such as smearing and Laplacian will use the other parameters. We call this a ``matrix'' because the fermion matrix inherits it.

Many pure gauge functions are bound to the gauge field, which is implemented by embedding a gauge matrix into the gauge field, and all operations are called through the gauge matrix. For example, we can apply HYP smearing~\cite{Hasenfratz:2001hp} to a gauge field by calling a member function of \mono{LatticeGauge}. APE smearing~\cite{APE:1987ehd}, stout smearing~\cite{Morningstar:2003gk}, HYP smearing, Wilson flow~\cite{Luscher:2010iy}, Symanzik flow~\cite{BMW:2012hcm}, Wuppertal smearing~\cite{Gusken:1989qx}, staggered phase, Laplacian, covariant derivative, and more are supported.

The fermion matrix inherits the gauge matrix and sets the necessary parameters for a Wilson/clover/HISQ action. The parameters are chosen to be ``general'' and provide fairly good, though not necessarily optimal, performance. Users can apply their settings by changing default values in \mono{general.py}. The multigrid-related parameters are similar to those in QUDA's test example, with the \mono{geo\_block\_size} parameter left open to users. After defining a fermion action, the corresponding matrix is constructed. We can apply the matrix to a fermion field or obtain the propagator by performing inversion on a source field. The default iterative solvers are BiCGStab for Wilson/clover and CG for HISQ.

A \mono{Multigrid} instance is included in the fermion matrix if the multigrid solver is enabled. Note that the multigrid solver is only supported with Wilson and clover fermion matrix. We construct it if it is not initialized, and then it can be passed across different fermion matrices. This allows generating a \mono{Multigrid} with a light quark mass and using it with slightly heavier quark masses, a common technique.

We have implemented fermion forces for the clover action and HISQ action, which are used in the HMC algorithm. 1-flavor and 2-flavor clover fermion actions with forces are defined. We also have Wilson and Symanzik gauge actions with corresponding gauge forces. It is possible to implement an HMC with any flavor of clover quarks using PyQUDA. Note that the derivative of stout smearing has not been implemented in PyQUDA, so the fermion action with the stout smearing cannot be used in the HMC algorithm at present. HISQ fermion actions of any flavor with mass preconditioning \cite{Clark:2006wp} can be used with PyQUDA. This can be achieved by passing the rational approximation parameters to the pseudofermion.

Additional explanation is needed for the anisotropic clover fermion action. QUDA does not support generating a clover field on an anisotropic lattice directly. We must apply a fake ``anisotropy'' to the gauge field, make QUDA think it's an isotropic lattice, and then QUDA will calculate the anisotropic clover term we need. Another parameter is required to obtain the extra ``anisotropy'', which is why we need an extra \mono{xi\_0} parameter to construct the anisotropic clover fermion matrix.

Code autocompletion is enabled through modern LSP or IDE when type hints are provided for variables, classes, and functions. While stub files for PyQUDA are included with the package, they must be manually updated if QUDA headers change, as there is no script to automatically generate them. Consider that a mismatch of type hints will not result in any runtime warnings or errors, manual updates are acceptable.

\section{Quark propagators and correlation functions\label{section:propag}}

\subsection{Formulism}

We have the pion meson interpolator $\mathcal{O}_\pi(\vec{p},t)=\sum_{\vec{p}}e^{i\vec{p}\cdot\vec{x}}\bar{d}(\vec{x},t)\gamma_5u(\vec{x},t)$. A 2-point correlation function of the pion meson can be written as
\begin{equation}\label{eq:two-point-pre}
  \begin{aligned}
      & \left\langle\mathcal{O}_\pi(\vec{p},t)\mathcal{O}_\pi^\dagger(\vec{p},t_0)\right\rangle                                                                                          \\
    = & \sum_{\vec{x},\vec{x}_0}e^{-i\vec{p}\cdot\vec{x}}e^{i\vec{p}\cdot\vec{x}_0}\mathrm{Tr}\left[S^d(\vec{x}_0,t_0;\vec{x},t)\gamma_5 S^u(\vec{x},t;\vec{x}_0,t_0)\gamma_5\right]     \\
    = & \sum_{\vec{x},\vec{x}_0,\vec{x}'_0}e^{-i\vec{p}\cdot\vec{x}}e^{i\vec{p}\cdot\vec{x}_0}\mathrm{Tr}\left[S^{d\dagger}(\vec{x},t;\vec{x}'_0,t_0)S^u(\vec{x},t;\vec{x}_0,t_0)\right]
  \end{aligned}
\end{equation}
The last equality in Eq.~(\ref{eq:two-point-pre}) uses the $\gamma_5$-Hermiticity: $\gamma S\gamma=S^\dagger$, and contains that gauge-variant values average to zero over the gauge ensemble. Since the $u$ and $d$ quarks are degenerate on the lattice, we have $S^u=S^d=S$. Let's explicitly write out the spin and color indices in the final equation:
\begin{equation}\label{eq:two-point}
  \begin{aligned}
    = & \sum_{\vec{x},\vec{x}_0,\vec{x}'_0,i,j,a,b}e^{-i\vec{p}\cdot\vec{x}}e^{i\vec{p}\cdot\vec{x}_0}S^*(\vec{x},t;\vec{x}'_0,t_0)^{ba}_{ji}S(\vec{x},t;\vec{x}_0,t_0)^{ba}_{ji} \\
    = & \sum_{\vec{x},i,j,a,b}e^{-i\vec{p}\cdot\vec{x}}S_{\vec{0},t_0}^*(\vec{x},t)^{ba}_{ji}S_{\vec{p},t_0}(\vec{x},t)^{ba}_{ji}
  \end{aligned}
\end{equation}
where $i,j$ are spin indices and $a,b$ are color indices. The last equality in the Eq.~(\ref{eq:two-point}) uses the definition $S_{\vec{0},t_0}(\vec{x},t)=\sum_{\vec{x}_0}S(\vec{x},t;\vec{x}_0,t_0)$ and $S_{\vec{p},t_0}(\vec{x},t)=\sum_{\vec{x}_0}e^{-i\vec{p}\cdot\vec{x}_0}S(\vec{x},t;\vec{x}_0,t_0)$, which are definitions of propagators from wall sources with the momentum $\vec{0}$ and $\vec{p}$.

Inserting a vector current in $z$ direction $V_3=\bar{u}\gamma_3 u$ between the two operators defines a 3-point correlation function,
\begin{equation}\label{eq:three-point}
  \begin{aligned}
      & \sum_{\vec{x}_1}\left\langle\mathcal{O}_\pi(\vec{p},t)V_3(\vec{x}_1,t_1)\mathcal{O}_\pi^\dagger(\vec{p},t_0)\right\rangle                      \\
    = & \sum_{\vec{x}_1,i,j,k,a,b}S^*_{\vec{p},t;\vec{0},t_0}(\vec{x}_1,t_1)^{ba}_{ji}(\gamma_5\gamma_3)_{jk} S_{\vec{p},t_0}(\vec{x}_1,t_1)^{ba}_{ki}
  \end{aligned}
\end{equation}
where
\begin{equation}\label{eq:sequential}
  \begin{aligned}
    S_{\vec{p},t;\vec{0},t_0}(\vec{x}_1,t_1)= & \sum_{\vec{x}}e^{-i\vec{p}\cdot\vec{x}}S(\vec{x}_1,t_1;\vec{x},t)\gamma_5 S_{\vec{0},t_0}(\vec{x},t) \\
    =                                         & \sum_{\vec{x}}S(\vec{x}_1,t_1;\vec{x},t)X_{\vec{p},t;\vec{0},t_0}(\vec{x},t)
  \end{aligned}
\end{equation}
$X_{\vec{p},t;\vec{0},t_0}$ is the sequential source. The required sequential propagator $S_{\vec{p},t;\vec{0},t_0}$ can be calculated by performing the inversion on the sequential source.

\subsection{Example}

In this section, we will demonstrate a practical lattice QCD calculation. We will generate quark propagators from wall sources with a zero and a non-zero momentum using the clover fermion matrix, and then contract them to obtain a 2-point correlation function with the powerful \mono{einsum} function. After generating the sequential propagator from the sequential source, a 3-point correlation function with the vector current is also constructed. The code for this process is shown in Listing~\ref{listing:propag}

\begin{widetext}
  \begin{lstlisting}[language=Python, caption={Example for quark propagators and correlation functions.}, label=listing:propag]
import cupy as cp
from pyquda import init
from pyquda_utils import core, io, phase, gamma

xi_0, nu = 2.464, 0.95
m_0 = 0.3
clover_csw_t, clover_csw_r = 0.91, 1.07
tol, maxiter = 1e-12, 100

init([1, 1, 1, 2], resource_path=".cache")
latt_info = core.LatticeInfo([4, 4, 4, 8], t_boundary=-1, anisotropy=xi_0 / nu)

gauge = io.readChromaQIOGauge("weak_field.lime")
exp_ipx = phase.MomentumPhase(latt_info).getPhase([0, 0, 1])
t0, t = 0, 4

dirac = core.getDirac(latt_info, m_0, tol, maxiter, xi_0, clover_csw_t, clover_csw_r)
dirac.loadGauge(gauge)
S_0 = core.invert(dirac, source_type="wall", t_srce=t0)
S_p = core.invert(dirac, source_type="wall", t_srce=t0, source_phase=exp_ipx)
X_p0 = core.LatticePropagator(S_0.latt_info)
X_p0.data[:, t] = cp.einsum(
    "wzyx,ij,wzyxjkab->wzyxikab",
    exp_ipx[:, t].conj(),
    gamma.gamma(15),
    S_0.data[:, t],
    optimize=True,
)
S_p0 = core.invertPropagator(dirac, X_p0)
dirac.destroy()
io.writeNPYPropagator("S_0.npy", S_0)

twopt = cp.einsum(
    "wtzyx,wtzyxjiba,wtzyxjiba->t",
    exp_ipx.conj(),
    S_0.data.conj(),
    S_p.data,
    optimize=True,
)
threept = cp.einsum(
    "wtzyxjiba,jk,wtzyxkiba->t",
    S_p0.data.conj(),
    gamma.gamma(15) @ gamma.gamma(4),
    S_p.data,
    optimize=True,
)
  \end{lstlisting}
\end{widetext}

The Python scripts often start with the \mono{import} statement. The package name for PyQUDA is \mono{pyquda}, and we need the \mono{init} function and the \mono{core} module from it. Additionally, I/O functions are required to read in a gauge field from a QIO file and to write out a propagator field to an NPY file. Next, we set some parameters for the fermion action from lines 5 to 8. The anisotropic clover fermion matrix is defined in the form of Eq.~(\ref{eq:clover}),
\begin{equation}\label{eq:clover}
  \begin{aligned}
    2\kappa a_t\mathcal{M}=1-\kappa & \left[\gamma_4W_4+\frac{\nu}{\xi_0}\sum_{i=1}^{3}\gamma_iW_i\right.                                                                \\
                                    & \left.+c_{SW}^t\sum_{i=1}^{3}\sigma_{i4}\hat{F}_{i4}+\frac{c_{SW}^r}{\xi_0}\sum_{i=1}^{3}\sum_{j>i}\sigma_{ij}\hat{F}_{ij}\right],
  \end{aligned}
\end{equation}
where $\kappa^{-1}=2(m_0+1+3\frac{\nu}{\xi_0})$. Parameters like \mono{m\_0}, \mono{xi\_0}, \mono{nu}, \mono{clover\_csw\_t} and \mono{clover\_csw\_r} are defined accordingly. \mono{tol} and \mono{maxiter} represent the tolerance and the maximum number of iterations for the solver, respectively.

We initialize the PyQUDA environment using the \mono{init} function in line 10. In this case, we use 2 processes to divide the lattice into two parts along the $t$ direction. The \mono{resource\_path} parameter functions similarly to the environment variable \mono{QUDA\_RESOURCE\_PATH}, indicating where to save the tuning cache files. CuPy is the default backend, meaning PyQUDA uses CuPy to handle data on the GPU memory. The backend can be changed by passing the corresponding string to the \mono{backend} parameter in \mono{init}: possible options include \mono{"numpy"}, \mono{"cupy"}, and \mono{"torch"}. In line 11, a lattice of size $4^3\times8$ is defined using \mono{LatticeInfo}, which implies that each GPU will handle a $4^4\times4$ segment. The lattice is anti-periodic in the $t$ direction and has an anisotropy defined by \mono{xi\_0 / nu}. These parameters determine how the gauge field should be prepared before passing it to QUDA. In lines 13 and 14, the gauge field is read from a lime file using \mono{readChromaQIOGauge}, and the phase with momentum mode \mono{[0, 0, 1]} is created using \mono{MomentumPhase}. The source time $t_0$ and the sink time $t$ (used in the 3-point situation) are set to 0 and 4 (the middle of the time direction).

Next, we set up the fermion matrix. The parameters defined above are passed to the \mono{getDirac} function in line 17 to obtain the clover fermion matrix, defined as $a_t\mathcal{M}$ in Eq.~(\ref{eq:clover}). Since it is not possible to do the computation for the matirx without knowing the gauge field, we use \mono{loadGauge} in line 18 to send the gauge field to QUDA, where the clover field is then created. We can enable the multigrid solver by passing the block size to \mono{getDirac}. For instance, \mono{core.getDirac(..., multigrid=[[2, 2, 2, 2], [2, 2, 2, 2]])} indicates the use of a 2-level multigrid solver, projecting a $2^4$ sublattice to a coarse lattice site at each level. When the block size is set, the multigrid solver setup will occur during the gauge field loading. The multigrid configuration can then be accessed via \mono{dirac.multigrid}, which can be passed to the \mono{multigrid} parameter in another \mono{getDirac} call like \mono{core.getDirac(..., multigrid=dirac.multigrid)}.

The most time-consuming part of the script is \mono{core.invert} and \mono{core.invertPropagator} in lines 19, 20 and 29. This function combines two tasks:
\begin{itemize}
  \item Creating a source field. The \mono{invert} function accepts \mono{source\_type} and \mono{t\_srce} parameters control the source's properties. For example, \mono{"wall"} and \mono{t0} indicate the source has non-zero values only at timeslice $t=t_0$, known as a wall source. If \mono{source\_phase} is not set, the values in the wall are 1; otherwise, the corresponding values in \mono{source\_phase} will be used to fill the wall, creating a momentum source for the second propagator. The \mono{invertPropagator} function accepts an existing propagator as the source field, where we pass the sequential source \mono{seq\_source} here.
  \item Performing the inversion on the source field. The propagator is a matrix in the spin-color space. The inversion, implemented by \mono{dirac.invert}, is currently performed on a spin-color scalar, so 12 inversions are needed to obtain the full propagator. The two functions simplify this process, allowing us to get the propagator within a single line. Future improvements may involve calling the \mono{invertMultiSrcQuda} function in QUDA.
\end{itemize}

Line 30 destroys the multigrid configuration if \mono{dirac} has one, and this should be done only once per configuration. The propagator is written to an NPY file by \mono{writeNPYPropagator} in line 31, which can be then read with PyQUDA's \mono{readNPYPropagator} or with \mono{numpy.load} with pure NumPy (without MPI support).

Finally, we need to contract the freshly generated propagators to obtain the correlation functions. We express the coordinates $\vec{x},t$ in the even-odd preconditioned form, using letter $w$ to represent the even-odd parity. The axis order $w,t,z,y,x$ is chosen to align with the field layout in PyQUDA. The contraction for 2-point correlation function in Eq.~(\ref{eq:two-point}) becomes
\begin{equation}\label{eq:two-point-contract}
  \sum_{w,x,y,z,i,j,a,b}e^{-i\vec{p}\cdot\vec{x}}S_{\vec{0},t_0}^*(w,t,z,y,x)^{ba}_{ji}S_{\vec{p},t_0}(w,t,z,y,x)^{ba}_{ji}
\end{equation}
and for 3-point correlation function in Eq.~(\ref{eq:three-point}) becomes
\begin{equation}\label{eq:three-point-contract}
  \sum_{w,x,y,z,i,j,k,a,b}S^*_{\vec{p},t;\vec{0},t_0}(w,t,z,y,x)^{ba}_{ji}(\gamma_5\gamma_3)_{jk} S_{\vec{p},t_0}(w,t,z,y,x)^{ba}_{ki}
\end{equation}
Now, we can translate these contractions into \mono{einsum} operations. The subscripts and operands in lines 33 to 39 of the code directly correspond to those in Eq.~(\ref{eq:two-point-contract}). Specifically:
\begin{itemize}
  \item \mono{exp\_ipx.conj()}:  the $e^{-i\vec{p}\cdot\vec{x}}$ term, with indices \mono{"wtzyx"};
  \item \mono{S\_0.data.conj()}: the $S_{\vec{0},t_0}^*$ term, with indices \mono{"wtzyxjiba"};
  \item \mono{S\_p.data}: the $S_{\vec{p},t_0}$ term, with indices \mono{"wtzyxjiba"}.
\end{itemize}
Similarly, lines 40 to 46 is the direct translation of Eq.~(\ref{eq:three-point-contract}):
\begin{itemize}
  \item \mono{S\_p0.data.conj()}: the $S_{\vec{p},t;\vec{0},t_0}^*$ term, with indices \mono{"wtzyxjiba"};
  \item \mono{gamma\_5 @ gamma\_3}: the $\gamma_5\gamma_3$ term, with indices \mono{"jk"};
  \item \mono{S\_p.data}: the $S_{\vec{p},t_0}$ term, with indices \mono{"wtzyxkiba"}.
\end{itemize}
The sequential propagator used in the 3-point correlation function defined in Eq.~(\ref{eq:sequential}) is generated in lines 21 to 29. The \mono{einsum} function call here is used to calculate the sequential source
\begin{equation}
  X_{\vec{p},t;\vec{0},t_0}(w,t,z,y,x)^{ab}_{ik}=e^{-i\vec{p}\cdot\vec{x}}(\gamma_5)_{ij} S_{\vec{0},t_0}(\vec{x},t)^{ab}_{jk}
\end{equation}
\begin{itemize}
  \item \mono{X\_p0.data}: the $X_{\vec{p},t;\vec{0},t_0}$ term, with indices \mono{"wtzyxikab"};
  \item \mono{exp\_ipx.conj()}: the $e^{-i\vec{p}\cdot\vec{x}}$ term, with indices \mono{"wtzyx"};
  \item \mono{gamma\_5}: the $\gamma_5$ term, with indices \mono{"ij"};
  \item \mono{S\_0.data}: the $S_{\vec{0},t_0}$ term, with indices \mono{"wtzyxjkab"}.
\end{itemize}
We pick the sink time $t=4$ in the code, and then we can suppress the \mono{"t"} index in the \mono{einsum} call.

Notice that we are using the \mono{data} member of \mono{LatticePropagator}s as inputs, which are CuPy NDArrays. This approach yields the 2-point correlation function for the pion meson with momentum mode \mono{[0, 0, 1]} from source time $t_0=0$ to all sink times $t$, and the 3-point correlation function for the pion meson with momentum mode \mono{[0, 0, 1]} from source time $t_0=0$ to sink time $t=4$ with a vector current inserted at all times $t_1$.

Once correlation functions for sublattices are obtained, we need to gather them on one process and calculate the correlation functions for the whole lattice. This can be done by MPI for Python package easily. After getting the correlation functions on an ensemble, one can leverage them with other Python packages, such as lsqfit for fitting physical observables, or Matplotlib for visualization, etc.

\section{Hybrid Monte Carlo\label{section:hmc}}

\subsection{Example}

The code for the HMC algorithm with the Symanzik gauge action and $N_f=2+1$ clover fermion action is shown in Listing~\ref{listing:hmc}.

\begin{widetext}
  \begin{lstlisting}[language=Python, caption={Example for hybrid Monte Carlo.}, label=listing:hmc]
from math import exp
from time import perf_counter

from pyquda import init
from pyquda.hmc import HMC, O4Nf5Ng0V
from pyquda.action import GaugeAction, CloverWilsonAction
from pyquda_utils import core, io
from pyquda_utils.hmc_param import symanzik_tree_gauge, wilson_rational_param

beta, u_0 = 4.8665, 0.949
mass_l, mass_s = 0.3, 0.5
clover_csw = 1.17
tol, maxiter = 1e-12, 1000
start, stop, warm, save = 0, 2000, 500, 5
t, n_steps = 1.0, 5

init(resource_path=".cache")
latt_info = core.LatticeInfo([4, 4, 4, 8], t_boundary=-1, anisotropy=1.0)

monomials = [
    GaugeAction(latt_info, symanzik_tree_gauge(u_0), beta, u_0),
    CloverWilsonAction(latt_info, wilson_rational_param[2], 0.3, 2, tol, maxiter, clover_csw),
    CloverWilsonAction(latt_info, wilson_rational_param[1], 0.5, 1, tol, maxiter, clover_csw),
]

hmc = HMC(latt_info, monomials, O4Nf5Ng0V(n_steps))
gauge = core.LatticeGauge(latt_info)
hmc.initialize(10086, gauge)

plaq = hmc.plaquette()
core.getLogger().info(f"Trajectory {start}:\n" f"Plaquette = {plaq}\n")
for i in range(start, stop):
    s = perf_counter()

    hmc.gaussMom()
    hmc.samplePhi()

    kinetic_old, potential_old = hmc.momAction(), hmc.gaugeAction() + hmc.fermionAction()
    energy_old = kinetic_old + potential_old

    hmc.integrate(t, 2e-15)

    kinetic, potential = hmc.momAction(), hmc.gaugeAction() + hmc.fermionAction()
    energy = kinetic + potential

    accept = hmc.accept(energy - energy_old)
    if accept or i < warm:
        hmc.saveGauge(gauge)
    else:
        hmc.loadGauge(gauge)

    plaq = hmc.plaquette()
    core.getLogger().info(
        f"Trajectory {i + 1}:\n"
        f"Plaquette = {plaq}\n"
        f"P_old = {potential_old}, K_old = {kinetic_old}\n"
        f"P = {potential}, K = {kinetic}\n"
        f"Delta_P = {potential - potential_old}, Delta_K = {kinetic - kinetic_old}\n"
        f"Delta_E = {energy - energy_old}\n"
        f"acceptance rate = {min(1, exp(energy_old - energy)) * 100:.2f}%\n"
        f"accept? {accept or i < warm}\n"
        f"HMC time = {perf_counter() - s:.3f} secs\n"
    )

    if (i + 1) % save == 0:
        io.writeNPYGauge(f"./DATA/cfg/cfg_{i + 1}.npy", gauge)

  \end{lstlisting}
\end{widetext}

We start by importing the required modules and functions, similar to the quark propagator script. Currently, the HMC algorithm can only be applied to an isotropic lattice, requiring just one clover coefficient as specified in line 12. Additional parameters in lines 14 and 15 control the Markov chain and the molecular dynamics (MD) integration, respectively. The Markov chain begins at step 0 and ends at step 2000, using the first 500 steps to reach equilibrium. PyQUDA is initialized with a single process, and then an isotropic lattice with size \mono{[4, 4, 4, 8]} is defined in line 18. Note that we import \mono{HMC} and an integrator named \mono{O4Nf5Ng0V}~\cite{OMELYAN2003272} in line 5, which wraps most of the operations in the HMC algorithm.

The action is defined in monomials in lines 20 to 24. The parameters used to set up these actions are similar to those in \mono{getDirac}, but the actions are defined within the \mono{pyquda.action} submodule. A \mono{HMC} object is created and initialized with a unit gauge field that has just been generated in line 27, indicating that the Markov chain begins with a unit gauge field. The number \mono{10086} passed to the initialization function is the seed for the random number generator. In a fermion action, we need to define the fermion matrix, the square root of the matrix, and the inverse of the matrix. These matrices might be constructed using rational approximation, and the corresponding coefficients are defined in \mono{wilson\_rational\_param}, and the item index 1 and 2 return coefficients for 1-flavor and 2-flavor, respectively. The integrator is initialized with the number of steps for the symplectic integration.

The main part of the HMC algorithm is implemented in the loop from \mono{start} to \mono{end}. Before delving into this part, let's discuss some HMC-algorithm-irrelevant code within the loop. A timer starts at the beginning of each step, and the time consumed is printed at the end, as shown in lines 33 and 63. Information about the current step is printed in lines 53 to 63, including the plaquette of the new gauge field and the acceptance rate of the step. The generated gauge fields are saved every \mono{save} steps, as specified in line 65 and 66.

Lines 34 to 50 detail how the HMC algorithm operates. First, we generate the conjugate momentum field from a Gaussian distribution, which can be easily accomplished using \mono{hmc.gaussMom}. Next, we generate the pseudo fermion field for each fermion action in the previously defined monomial list. This involves generating a fermion field from a Gaussian distribution and then applying the square root of the fermion matrix in the action to it. The \mono{hmc.samplePhi} function is a wrapper for these operations and should work well if we have defined the square root.

The Hamiltonian should be calculated before the MD integration. In line 38, we compute the potential energy from the gauge field and the kinetic energy from the momentum field; the Hamiltonian is their sum. Line 41 performs the integration using the \mono{O4Nf5Ng0V} integrator mentioned earlier, evolving the system by 1.0 units of evolution time. The integration is divided into 5 parts to reduce Hamiltonian errors. Due to the truncation error of floating point numbers, a projection to SU(3) is performed on the new gauge field, and the prejection precision is passed to the function. Another Hamiltonian calculation in line 43 yields the new potential and kinetic energy of the field, allowing us to compute the $\Delta\mathcal{H}$ for the step.

According to the Metropolis algorithm~\cite{10.1063/1.1699114}, the acceptance rate is the minimum of 1 and $e^{-\Delta\mathcal{H}}$ , which determines whether to accept the new gauge field in line 46. If the gauge field is accepted (or if the Markov chain is still warming up), the gauge field is saved to \mono{gauge} using \mono{saveGauge} in line 48; otherwise, the gauge field is restored to the previous state using \mono{loadGauge} in line 50.

Finally we have another example to show how sub integrator works. The symplectic integration in the MD process is wrapped as \mono{hmc.integrate} in line 41, which uses the \mono{O4Nf5Ng0V(n\_steps)} algorithm. The integrator is initialized with \mono{n\_steps=5} and it will calculate the force term 26 ($5 \times 5 + 1$) times in one step, which is time-consuming. An alternative trajectory is using lower order integrator for fermion force and higher order integrator for gauge force. The example to generate such a \mono{HMC} object is shown in Listing~\ref{listing:subintegrator}.

\begin{widetext}
  \begin{lstlisting}[language=Python, caption={Example for sub integrator.}, label=listing:subintegrator]
hmc_inner = HMC(latt_info, monomials[:1], O4Nf5Ng0V(4))
hmc = HMC(latt_info, monomials[1:], O4Nf5Ng0V(3), hmc_inner)
  \end{lstlisting}
\end{widetext}

Consider that the time to compute the gauge force term is much less than that of the fermion force term, we use a \mono{O4Nf5Ng0V(3)} to integrate the fermion actions and a \mono{O4Nf5Ng0V(12)} to integrate the gauge action, which computes the fermion force term 16 times in every step. It is faster than integrating all the actions with \mono{O4Nf5Ng0V(5)}, and get a silimar $\Delta\mathcal{H}$.

\section{Summary and outlook\label{section:summary}}

We have developed a Python wrapper for the QUDA library to accelerate lattice QCD calculations in Python. This project have already been used in our works~\cite{Shi:2023sdy,Shi:2024fyv,Li:2024pfg,Li:2023zig}. By translating QUDA C APIs to Cython code, we can call QUDA functions like smearing and fermion matrix inversion easily from Python, and it becomes nature to use other powerful Python packages to process and analyze the data.

We want to add more features to the PyQUDA project, and we list some of them which are in high priority blow:
\begin{itemize}
  \item Design an interface to implement the plugin system, which enables calling C/C++/CUDA functions from Python for specific measurement like baryon correlation function.
  \item Automatically generate the rational parameter for HISQ fermion action in the HMC algorithm. A Python script which is the translation of AlgRemez\footnote{{maddyscientist/AlgRemez: Implementation of the Remez algorithm} \url{https://github.com/CLQCD/PyQUDA}} is used for now, and we need to embed it into PyQUDA.
  \item Implement the derivative for the stout smearing. Adding stout smearing to a fermion action is widely used in recent lattice QCD researches~\cite{CLQCD:2023sdb,Du:2024wtr} and we need to compute the force term for the smearing. This is not implemented yet in QUDA and we need to add such a kernel and then expose this to the C API.
  \item Implement the Hasenbusch clover action~\cite{Hasenbusch:2001ne} or something similar to speedup the HMC algorithm. Modifications to QUDA are needed for this feature.
\end{itemize}

\begin{acknowledgments}
  This work is based on the QUDA library~\cite{Clark:2009wm,Babich:2011np,Clark:2016rdz}. This work is supported in part by NSFC grants No. 12293060, 12293062, 12293065, 12047503 and 11935017, the science and education integration young faculty project of University of Chinese Academy of Sciences, the Strategic Priority Research Program of Chinese Academy of Sciences, Grant No. XDB34030303 and YSBR-101. The numerical calculations in this work were carried out on the ORISE Supercomputer.
\end{acknowledgments}

\bibliography{ref}

\end{document}